# Optimized Pap Smear Image Enhancement: Hybrid PMD Filter-CLAHE Using Spider Monkey Optimization

Ach Khozaimi [1,2], Isnani Darti [1], Syaiful Anam [1], Wuryansari Muharini Kusumawinahyu [1]

[1]Department of Mathematics, Faculty of Mathematics and Natural Sciences, Brawijaya University, Malang, Indonesia
[2]Department of Computer Science, Faculty of Engineering, Trunojoyo University of Madura, Indonesia

**ABSTRACT**

Pap smear image quality is crucial for cervical cancer detection. This study introduces an optimized hybrid approach that combines the Perona-Malik Diffusion (PMD) filter with contrast-limited adaptive histogram equalization (CLAHE) to enhance Pap smear image quality. The PMD filter reduces the image noise, whereas CLAHE improves the image contrast. The hybrid method was optimized using spider monkey optimization (SMO PMD-CLAHE). BRISQUE and CEIQ are the new objective functions for the PMD filter and CLAHE optimization, respectively. The simulations were conducted using the SIPaKMeD dataset. The results indicate that SMO outperforms state-of-the-art methods in optimizing the PMD filter and CLAHE. The proposed method achieved an average effective measure of enhancement (EME) of 5.45, root mean square (RMS) contrast of 60.45, Michelson's contrast (MC) of 0.995, and entropy of 6.80. This approach offers a new perspective for improving Pap smear image quality.

*Corresponding Author:*

Isnani Darti
Department of Mathematics, Faculty of Mathematics and Natural Sciences, Brawijaya University, Malang, Indonesia
Email: isnanidarti@ub.ac.id

## 1. INTRODUCTION

Cancer remains a global health concern, and cervical cancer is a substantial health hazard, particularly in developing countries [1]. In 2020, Indonesia reported 36,633 cases of cervical cancer, trailing only for breast cancer [2]. The Global Cancer Observatory predicted 570,000 new cases and 311,000 deaths globally in 2018 [3]. Approximately 90% of cervical cancer-related deaths occur in developing countries [4]. Efforts to reduce cervical cancer mortality include incorporating information technology and artificial intelligence into screening procedures [5]. Noise reduction and contrast enhancement improve medical image quality before classification. Several studies have shown that reducing image noise substantially increases classification accuracy [6][7]. The PMD filter reduces image noise and smoothens images while preserving essential edges [8]. PMD uses a modified Gaussian function to weigh each pixel value, with higher values at the center and lower values at the periphery [9]. Studies have shown that PMD filters extract and identify malignant tumors in medical images. [10]. In addition, the PMD filter has improved the deep learning performance in cervical cancer classification.[11]. However, the PMD filter performance relies on the fine-tuning of its parameters. Tsiotsios and Petrou (2013) chose the PMD filter parameter iteratively [12]. The other research uses Particle Swarm Optimization (PSO) to select PMD parameters and improve the PMD filter performance [13].

CLAHE is a variant of adaptive histogram equalization (AHE) that limits contrast enhancement [14]. CLAHE effectively improves Pap smear images and enhances VGG16, InceptionV3, and Efficient Net performance in cervical cancer classification [15]. CLAHE has successfully improved image quality and enhanced the performance of various machine learning algorithms such as K-Nearest Neighbors (KNN) and Artificial Neural Networks (ANN) in cervical cancer classification [16]. In other research, CLAHE enhances the performance of the You Only Look Once (YOLO) algorithm in recognizing road markings at night [17].

CLAHE has also improved CNN performance in the CT scan image segmentation of lung cancer [18] and water image classification [19]. However, the effectiveness of CLAHE depends on its parameters, i.e., clip limit and tile size. Qassim H. et al. set up a clip limit of 0.01 and a tile size of 8 × 8 to get the best CLAHE performance enhancing dental digital X-ray images [20]. Several studies have applied different heuristic optimization algorithms to improve CLAHE performance.

PSO was used to optimize CLAHE performance with multi-objective functions, i.e., entropy and SSIM. This approach maximized image contrast while minimizing distortion in X-ray medical images [21]. Fawzi et al. applied the Whale Optimization Algorithm (WOA) to optimize CLAHE performance with DataSignal as the objective function. The DataSignal results from multiplying the Entropy by the peak signal-to-noise ratio (PSNR). It effectively enhances image contrast across datasets like faces-1999, BraTS, and Pasadena-houses 2000 [22]. Kuran and Kuran, 2022, employed the Cuckoo Search Algorithm (CSA) with entropy and Fast Noise Variance Estimation (FNVE) as objective functions. This study showed superior performance in CLAHE optimization on the CEED2016 dataset compared to the Bat Firefly and Flower Pollination Algorithms (FPA) [23]. In 2022, FPA optimized CLAHE with entropy and FNVE as objective functions. This study achieved notable noise reduction and contrast enhancement on Pasadena-houses 2000 and DIARETDB0 datasets [24]. Surya and Muthukumaravel, 2023, used Adaptive Sailfish Optimization (ASFO) to enhance CLAHE performance. This study focuses on maximizing contrast and entropy with successful enhancement outcomes on mammogram images from the MIAS database [25]. Cat Swarm Optimization (CSO) is also used to enhance CLAHE performance with entropy and FNVE as objective functions. This approach outperformed traditional methods like HSL, EC, HE, and CLAHE-CSA on the CEED2016 dataset [26]. In 2024, Haddadi et al. introduced the Pelican Optimization Algorithm (POA) to optimize CLAHE performance with several metrics, including PSNR, mean squared error (MSE), entropy, and structure similarity index measure (SSIM) as objective functions. This study uses a private dataset and outperformed the existing image enhancement techniques [27]

In this study, we aimed to enhance Pap smear image quality using a hybrid PMD filter - CLAHE. The PMD filter is used for noise reduction, and CLAHE is used for contrast enhancement. The hybrid PMD–CLAHE was optimized using the Spider Monkey Optimization (SMO) algorithm. SMO performs best in optimizing UCAV path-planning problems compared to other metaheuristic algorithms [28]. A new objective function was introduced in this study. The blind/reference-less image spatial quality evaluator (BRISQUE) is a new objective function for PMD filter optimization. BRISQUE is highly competitive with this no-reference image quality assessment (NR-IQA) approach. It is also statistically better than the popular full-reference image quality assessment (FR-IQA), such as PSNR and SSIM [29]. Contrast enhancement-based image quality (CEIQ) is a new objective function for CLAHE optimization. The CEIQ is calculated based on the histogram's features of entropy, cross-entropy, and SSIM [30]. CEIQ determines that the enhanced image has contrast deformation [31].

This study used several metrics to evaluate the image denoising and contrast enhancement. MSE, SSIM, PSNR, CEIQ, entropy, enhancement measure estimation (EME), Michelson contrast (MC), and root mean square (RMS) contrast were used. These metrics evaluate image clarity, detail preservation, and contrast improvement. The proposed approach operates in the CIELAB color model of Pap smear images and offers several contributions. First, Hybrid SMO PMD-CLAHE provides the advantages of reducing noise and increasing contrast because most Pap smear images are noisy and have low contrast [32]. Second, BRISQUE and CEIQ are the new objective functions for the PMD filter and CLAHE optimization, respectively. BRISQUE was statistically better than PSNR and SSIM [29], while image contrast deformation can be evaluated by CEIQ [31]. Third, the SMO-PMD filter and SMO CLAHE outperformed state-of-the-art methods. This study offers a new perspective for improving Pap smear image quality and contributes to more accurate cervical cancer detection.

2. **METHOD**

Figure 1 shows the process for enhancing Pap smear images using a hybrid PMD filter and CLAHE techniques. The PMD filter and CLAHE were optimized using the SMO algorithm. The input was a color image obtained from the SIPaKMeD dataset. The SipakMed dataset contains 4,049 annotated cervical cell images, categorized into five classes: normal superficial squamous epithelial cells, normal intermediate squamous epithelial cells, normal columnar epithelial cells, low-grade squamous intraepithelial lesion (LSIL) cells, and high-grade squamous intraepithelial lesion (HSIL) cells. These images are captured using Pap smear techniques, emphasizing cervical cancer diagnosis. Each class represents distinct morphological features vital for medical classification. Its primary characteristics include high variability in cell shapes, textures, noise, and contrast levels. This variability poses challenges in accurate classification [33]. The color image is split into L (lightness), A (green-red), and B (blue-yellow) color channels in the CIELAB color space. The CIELAB color space is designed to resemble the human visual system (HVS) [34].

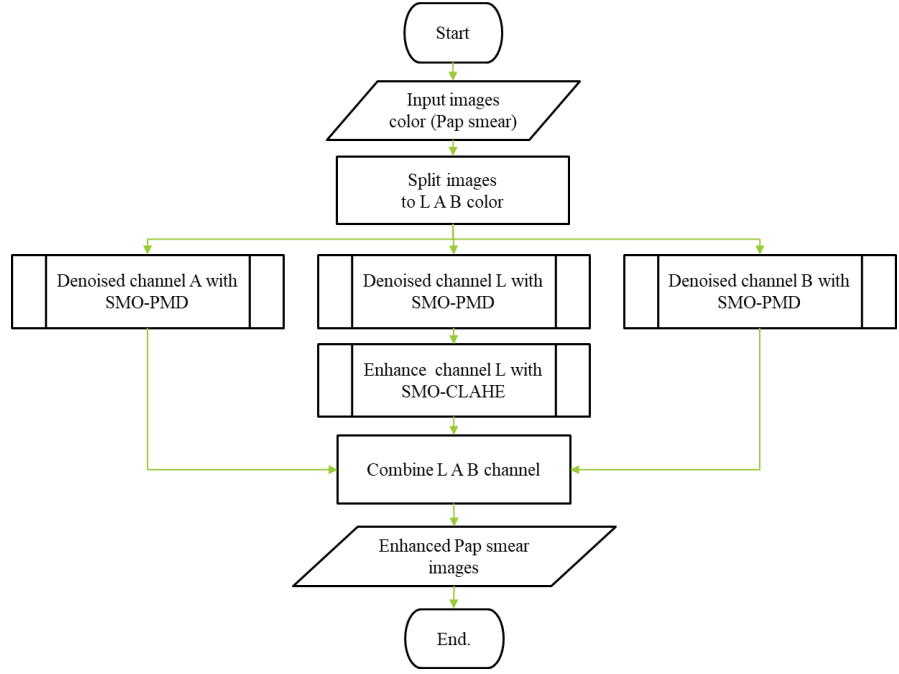

Figure 1. Flowchart of Pap smear image enhancement using SMO PMD-CLAHE

Each channel underwent separate processing steps.
1. Denoising: The A, B, and L channels are individually denoised using the SMO-PMD filter, which aims to reduce noise while preserving important image features such as edges.
2. Contrast Enhancement: After denoising, the lightness (L) channel was further enhanced using SMO-CLAHE, which improved the local contrast and highlighted finer details.

Once all channels (L, A, and B) were processed (denoised and contrast enhancement), they were recombined into the final enhanced Pap smear image. This enhanced image should exhibit an improved visual quality, reduced noise, and better contrast.

The hybrid PMD-CLAHE process was optimized using the SMO algorithm. The SMO optimizer was configured with 10 iterations and a population size of 50 to balance exploration and computational efficiency. The number of iterations (Niter) was set between 5 and 30 to control the degree of denoising. The diffusion coefficient ($\kappa$) ranged from 10 to 100 to adjust the smoothing intensity, while the gradient threshold ($\lambda$) was set between 0.1 and 0.25 to preserve image edges. The clip limit was set between 0.01 and 4 to manage contrast enhancement, and the tile size ranged from 2 to 16 to determine the local contrast regions. This configuration effectively balances the noise reduction and contrast enhancement.

### 2.1. Perona malik diffusion (PMD) filter

A PMD filter was used to reduce image noise while preserving the edges. This anisotropic diffusion process modifies the diffusion coefficient based on the gradient of the image, thereby enabling edge-preserving smoothing [9]. Given an image $I(x, y, t)$, where $x$ and $y$ are spatial coordinates and $t$ is the diffusion time (or iteration). The evolution of the image under the PMD was governed by the partial differential in Eq. (1).

$$\frac{\partial I}{\partial t} = \nabla \cdot (c(\| \nabla I \|)\nabla I), \qquad (1)$$

where $\frac{\partial I}{\partial t}$ represents the change in pixel intensity over time. $c(\| \nabla I \|)$ is the diffusion coefficient, which controls the amount of diffusion based on the gradient magnitude. The critical aspect of the Perona-Malik model is the choice of diffusion $c(\| \nabla I \|)$. The two common forms of diffusion coefficients are exponential and inverse quadratic [35]. This study used the exponential form of Eq. (2).

$$c(\| \nabla I \|) = e^{-\left(\frac{|\nabla I|}{K}\right)^2} \qquad (2)$$

where $K$ is a parameter that controls the sensitivity of the diffusion process to edges. Small values of $K$ result in more aggressive edge preservation, while larger values allow more smoothing. This iterative process is performed until a convergence condition is reached or a certain number of iterations is determined [36]. The edge-preserving property of the PMD filter comes from the behavior of the diffusion coefficient $c(\| \nabla I \|)$. Noise reduction is desired without blurring critical structural features, such as edges [37].

**2.2. Contrast-limited adaptive histogram equalization (CLAHE)**

CLAHE enhances image contrast in image processing. It divides the image into small regions, applies Histogram Equalization (HE) to each region using Eq. (4), and then redistributes the contrast by limiting the gain. This technique addresses the issue of excessive gain in homogeneous areas and is particularly beneficial in image processing because it enhances details and prevents the formation of artifacts [38]. The new grayscale value $(K_i)$ of the histogram equalization can be obtained by applying Eq. (3).

$$K_i = round\left(\frac{C_i(2^k-1)}{w\,h}\right) \tag{3}$$

where $C_i$ is the cumulative distribution of the $i$ grayscale value of the original image, $k$ is color variations in the image, $w$ and $h$ are image width and height. In the CLAHE technique, two variables control the contrast quality of the resulting image. Tile size is used to divide the image. The clip limit value $\beta$ is calculated as follows

$$\beta = \frac{P}{Q}\left(1 + \frac{\alpha}{100}(S_{max} - 1)\right), \tag{4}$$

where $P$ represent the area size and $Q$ is the total number of gray-level pixels in each block size (0-255), $S_{max}$ represents the maximum slope allowed in the histogram's cumulative distribution function (CDF) and $\alpha$ is the clip factor with range value (0 – 100). A clip limit prevents artifact creation by limiting noise amplification [39].

**2.3. Spider monkey optimization (SMO)**

The SMO algorithm is a global optimization method inspired by the social behavior of spider monkeys during foraging and exploration. SMO seeks an optimal solution to complex optimization problems by mimicking spider monkeys' collaborative and adaptive behaviors. It iterates through six phases: the Local Leader (LL) phase, the Global Leader (GL) phase, the Local Leader Learning (LLL) phase, the Global Leader Learning (GLL) phase, the Local Leader Decision (LLD) phase, and the Global Leader Decision (GLD) phase [40].

In SMO, each spider monkey in a group is represented as $SM_k(k = 1,2,...,N)$, serves as a potential solution. Each position vector of $SM_k$ in a $D$-dimensional space represents possible solutions, initialized using:

$$SM_{k,j} = SM_{min,j} + R\left(SM_{max,j} - SM_{min,j}\right), \tag{5}$$

$R$ is a random value between 0 and 1, and $SM_{max}$ and $SM_{min}$ upper and lower bounds are for each dimension. In the LL phase, each monkey's position is updated based on the local leader's guidance:

$$SM_{new\,i,j} = SM_{i,j} + R\left(Leader_{k,j} - SM_{i,j}\right) + U\left(SM_{r,j} - SM_{i,j}\right), \tag{6}$$

where $Leader_{k,j}$ is the local leader, $r$ is a randomly selected group member, and $U$ is a uniform random variable in the range $[-1,1]$. If the new position improves the solution, it is accepted; otherwise, it is discarded.

The GL phase updates positions based on global leadership, where the probability $prob_i$ is calculated using Eq. (7).

$$prob_i = 0.9 \times \frac{fit_i}{max\_fit} + 0.1 \tag{7}$$

with the highest-fitness monkey serving as the global leader. After each iteration, leaders are updated through greedy selection in the $L$ and $GL$ phases. The LLD phase prevents local leaders from stagnation by enforcing random position updates if a threshold (LocalLeaderLimit) is reached. Similarly, the GLD phase splits the group if the GlobalLeaderLimit threshold is met, thus encouraging further exploration.

**2.4. Image quality assessment (IQA)**

IQA is the process of assessing or evaluating the quality of a digital image. Three IQA models can be used: reduced reference image quality assessment (RR-IQA), FR-IQA, and NR-IQA [41]. This study used

MSE, SSIM, and PSNR to evaluate image denoising [42]. CEIQ, practical measure of enhancement measure estimation (EME), Michelson's contrast (MC), root mean square (RMS) contrast, and entropy are also used to evaluate image contrast enhancement.

EME is applied to quantify contrast-image enhancement, particularly for local contrast. It was calculated by dividing the image into blocks and considering the logarithmic ratio of the maximum and minimum intensities within each block.

$$EME = \frac{1}{M \times N} \sum_{i=1}^{M} \sum_{j=1}^{N} 20 \log \left( \frac{I_{max}(i,j)}{I_{min}(i,j)} \right) \qquad (8)$$

$M$ and $N$ are the number of blocks in the vertical and horizontal directions, respectively. $(I_{max}(i,j))$ and $(I_{min}(i,j))$ the maximum and minimum pixel intensities in the $i$ and $j$ block of the image. The logarithmic term helps measure contrast enhancement [43].

Michelson's Contrast (MC) is a simple contrast measure defined as the difference between an image's maximum and minimum intensity, divided by their sum $(I_{max})$ and $(I_{min})$ are the image's maximum and minimum pixel intensity [43].

$$MC = \frac{I_{max} - I_{min}}{I_{max} + I_{min}} \qquad (9)$$

The RMS contrast measures the overall contrast in an image by calculating the standard deviation of pixel intensities. $I(i,j)$ is the intensity at the pixel location $(i,j)$. $\bar{I}$ is the mean intensity of the entire image, and $M$ and $N$ are the image dimensions. The RMS contrast provides a single number that represents the contrast in an image, considering the variability in intensity values [43].

$$RMS\ Contrast = \sqrt{\frac{1}{MN} \sum_{i=1}^{M} \sum_{j=1}^{N} (I(i,j) - \bar{I})^2} \qquad (10)$$

The entropy measures the amount of information or randomness in an image. It is often used to assess texture or complexity.

$$Entropy = - \sum_{i=0}^{L-1} p_i \log_2(p_i) \qquad (11)$$

$L$ is the total number of possible intensity levels. $p_i$ is the probability (normalized histogram) of the occurrence of intensity level $(i)$. The entropy values range from 0 to $\log_2(L)$, with higher values indicating more complexity and randomness in the image [20].

The coefficient of Correlation (CoC) measures the correlation between pixel intensities in an original image and a processed image. A high correlation indicates that the processed image retains the structural information of the original. CoC determines how well image enhancement preserves the original structural details

$$CoC = \frac{\sum (I_x - \mu_x)(I_y - \mu_y)}{\sqrt{\sum (I_x - \mu_x)^2 \cdot \sum (I_y - \mu_y)^2}} \qquad (12)$$

$I_x$ and $I_y$ are pixel intensities in the original and enhanced images. $\mu_x$ and $\mu_y$ are mean intensities of the original and enhanced images.

Standard deviation (Std-dev) measures the spread of intensity values around the mean, reflecting the contrast variability in the image. Std-dev Quantify intensity variation and contrast.

$$Std\text{-}dev = \sqrt{\frac{1}{N} \sum_{i=1}^{N} (I_i - \mu)^2} \qquad (13)$$

$N$ is the total pixels in the image. $I_i$ is the intensity of the $i$-th pixel. $\mu$ is the mean intensity of the image.

### 2.5. Contrast enhancement-based image quality (CEIQ)

CEIQ is an image quality assessment technique that leverages contrast enhancement for evaluation [30]. This method employs histogram equalization to analyze and quantify image contrast. This process involves dividing the image histogram into multiple bins and calculating the average intensity value for each bin. Subsequently, these average values assign new intensity values to pixels within each corresponding bin. Figure 2 Shows the CEIQ evaluation model. CEIQ has two aspects of image quality assessment.
1. **The image similarity** measures the similarity of the original image to that of the contrast-enhanced image. The *Image similarity* was SSIM.
2. **Histogram entropy and cross-entropy** measure an even distribution of the image histogram. The entropy ($E$) equation is defined as Eq. (11)
Cross-entropy ($E_{xy}$) can be performed using the *histogram equalization* method. The cross-entropy values were calculated using Eq. (14)

$$E_{x,y} = -\sum_{i=0}^{b} h_x(i) \log h_y(i) \qquad (14)$$

$h_x$ is the histogram of the original image and $h_y$ the histogram of the contrast-enhanced image.

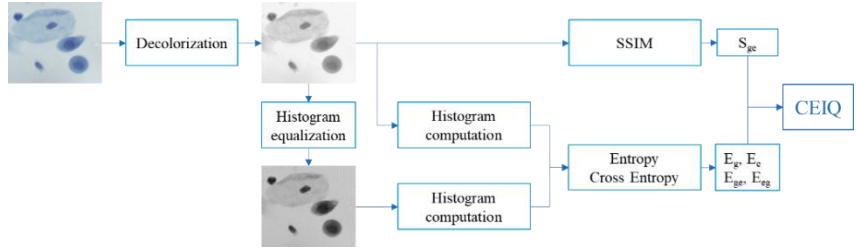

Figure 2. CEIQ Evaluation Model

### 2.6. Blind/ reference-less image spatial quality evaluator (BRISQUE)

BRISQUE is a model that calculates features directly from image pixels, unlike other methods that rely on transformations to different spaces, such as wavelets or Discrete Cosine Transformations [29]. Its efficiency does not require these transformations to extract features. BRISQUE assesses the image quality by comparing the input image to a model trained on images with similar distortions. It is trained on a database of natural scene images with known distortions and incorporates subjective quality scores, making it opinion-aware. Lower BRISQUE values indicate better perceptual image quality [29].

### 3. RESULTS AND DISCUSSION

This section presents the performance results of the SMO-PMD filter, SMO-CLAHE, and hybrid optimization of the PMD filter and CLAHE with SMO, referred to as SMO-PMD-CLAHE. Each method was specifically optimized to improve image-quality metrics for effective noise reduction, improved contrast, and enhanced image clarity.

### 3.1. SMO-PMD Filter

Table 1 shows the PMD filter optimization simulation results using PSO and SMO on ten images from SIPaKMeD. Overall, the SMO optimizer demonstrated superior performance across several key metrics compared with the PSO optimizer. Regarding MSE, SMO achieved a lower average error of 0.0456 compared with 0.0572 for PSO, indicating that SMO is more effective in optimizing the PMD filter and minimizing the error between the original and denoised images. Similarly, SSIM is slightly higher for SMO (0.9984 compared to 0.9981 for PSO), suggesting that SMO produces images with structural quality that closely resemble the original images. Regarding PSNR, SMO again outperforms PSO in optimizing the PMD filter, with an average of 62.26 dB, indicating that SMO yields images with less noise, whereas PSO's average is 61.00 dB. Both methods exhibited nearly identical entropy values, indicating that the informational content and details within the images were well-preserved in both cases. In the BRISQUE score as an objective function, SMO produces a slightly lower value (36.8561) than PSO (37.3073), signifying that SMO provides a marginally better subjective visual quality.

Table 1. Result simulation in optimizing PMD filter using PSO and SMO

| Images | Methods | MSE | SSIM | PSNR | Entropy | BRISQUE |
|---|---|---|---|---|---|---|
| 013_02 | PSO | 0.043947645 | 0.996258439 | 61.70144756 | 4.571843131 | 0.686698775 |
|  | **SMO** | **0.043389455** | **0.996305898** | **61.75696163** | **4.571857288** | **0.644838024** |
| 018_03 | PSO | 0.058383599 | 0.998492275 | 60.46789499 | 5.311107028 | 59.56556031 |
|  | **SMO** | **0.025031866** | **0.999318272** | **64.14587134** | **5.311234326** | **57.95722145** |
| 019_01 | PSO | **0.126389165** | **0.996250434** | **57.11370518** | 5.498297928 | 54.46574524 |
|  | SMO | 0.126785392 | 0.996229567 | 57.10011145 | 5.498286588 | **54.30521213** |
| 020_06 | PSO | 0.071592774 | 0.997519753 | 59.58211168 | **4.918457352** | 16.16094189 |
|  | SMO | **0.071446865** | **0.99752755** | **59.59097187** | 4.918444596 | **16.14697568** |
| 023_01 | PSO | 0.053882471 | 0.998797808 | 60.81632858 | 5.705432895 | 82.87267047 |
|  | SMO | **0.04483273** | **0.998983647** | **61.61485178** | 5.705490005 | **82.28349823** |
| 029_01 | PSO | 0.027944359 | 0.999283549 | 63.66786213 | 5.770480683 | 38.16081019 |
|  | SMO | **0.026237661** | **0.999327682** | **63.94155247** | 5.770483568 | **38.09620783** |
| 039_01 | PSO | 0.035305231 | 0.99859665 | 62.65241301 | **5.430661391** | 33.79115112 |
|  | SMO | **0.019000417** | **0.999242818** | **65.34317228** | 5.430631534 | **33.04454181** |
| 043_01 | PSO | 0.071355015 | 0.998425385 | 59.59655859 | 5.960775239 | 49.34384299 |
|  | SMO | **0.039904779** | **0.999094202** | **62.12055451** | **5.96146055** | **48.20374569** |
| 048_01 | PSO | 0.056730497 | 0.999230125 | 60.59263776 | 6.316787802 | 34.89882012 |
|  | SMO | **0.035746326** | **0.999523087** | **62.59848949** | 6.316873752 | **34.85028191** |
| 050_06 | PSO | 0.026868811 | 0.998198142 | 63.83831907 | 4.766578449 | 3.126566942 |
|  | SMO | **0.023824182** | **0.998404514** | **64.36062355** | 4.766553067 | **3.028702503** |
| Average | PSO | 0.057239957 | 0.998105256 | 61.00292786 | 5.42504219 | 37.3072808 |
|  | SMO | **0.045619967** | **0.998395724** | **62.25731604** | 5.425131527 | **36.85612253** |

These results suggest that SMO generally delivers a better image quality than PSO when optimizing the PMD filter. SMO consistently outperforms PSO in critical metrics, such as MSE, SSIM, PSNR, and BRISQUE. SMO-PMD filter offers new insight for applications requiring high image processing accuracy. Although the entropy values are similar between the two methods, SMO's consistent superiority in reducing error and noise.

### 3.2. SMO-CLAHE

The simulation results for CLAHE optimization using the POA and SMO algorithm on 10 images from the SIPaKMeD dataset can be seen in Figure 3. These results show relatively small differences across key metrics such as entropy, EME, RMS contrast, CoC, Standard deviation (STD-DEV), CEIQ, and processing time. Regarding entropy, the results were almost identical for both methods across all images, suggesting that the POA and SMO maintained similar levels of pixel intensity information. A similar trend is observed in the EME and RMS contrasts, where there is no significant difference between the two methods, indicating that both handle contrast enhancement similarly. One of the primary differences between the two methods is the processing time. SMO consistently outperformed POA in terms of speed. The average processing time for SMO was 7.5470 s, compared with 7.7650 s. This highlights the efficiency of SMO in terms of computational time, making it preferable in scenarios in which rapid image processing is essential, particularly for large-scale image datasets. The simulation results provide valuable insights into the performance of the POA and SMO in optimizing CLAHE on cervical images. Both methods showed comparable results in maintaining the image quality, as reflected in the near-identical values of entropy, EME, and RMS contrast. These metrics confirm that both POA and SMO can effectively enhance the contrast without significant loss of information. However, for practical implementation, processing time is a crucial factor. Therefore, SMO-CLAHE was more effective for cervical cancer detection.

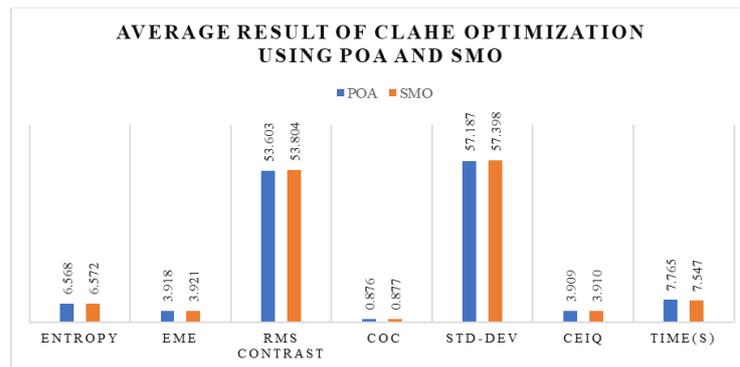

Figure 3. Average results of CLAHE optimization using POA and SMO

### 3.2.1. Hybrid SMO PMD-CLAHE

The average results for each evaluation metric on 10 Pap smear images using the SMO-PMD, SMO-CLAHE, and hybrid SMO PMD-CLAHE algorithms can be seen in Figure 4. The metrics used in this evaluation are Enhancement Measure Estimation (EME), Michelson Contrast (MC), RMS Contrast, Entropy, and CEIQ (Contrast-based Enhancement Image Quality). The SMO PMD method achieved the lowest EME value of 1.23, indicating limited effectiveness in enhancing illumination quality. SMO CLAHE demonstrated a significant improvement with an EME value of 3.85, while the combination of SMO PMD-CLAHE achieved the highest value of 5.45. This confirms that combining PMD and CLAHE has a synergistic effect, resulting in images with superior illumination quality. For Michelson Contrast, SMO PMD and SMO PMD-CLAHE achieved nearly optimal values of 1.00 and 0.99, respectively, indicating excellent contrast distribution. On the other hand, SMO CLAHE produced a lower MC value of 0.85, indicating slightly reduced contrast compared to the different methods.

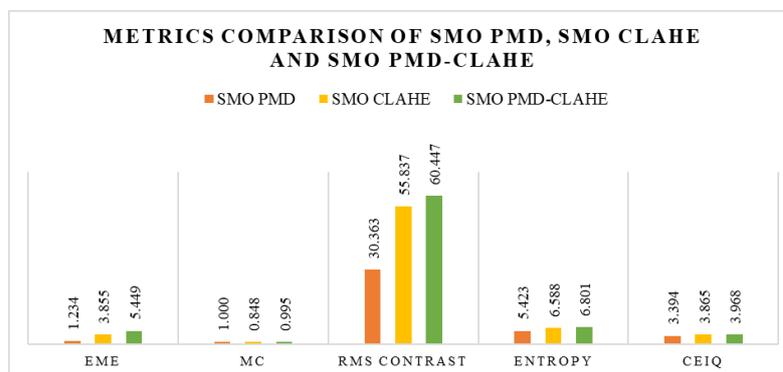

Figure 4. The average result on PMD, CLAHE, and Hybrid PMD-CLAHE Optimization using SMO

The SMO PMD method had the lowest RMS Contrast value of 30.36, suggesting limited enhancement capability. In contrast, SMO CLAHE showed a significant improvement with a value of 55.83, while SMO PMD-CLAHE achieved the highest value of 60.45. This demonstrates that combining PMD and CLAHE provides richer and more optimal contrast in the resulting images. The Entropy values reflect the diversity of information in the images. SMO PMD recorded the lowest value of 5.42, indicating less detailed images. SMO CLAHE achieved a higher entropy value of 6.59. At the same time, the combination of SMO PMD-CLAHE excelled with the highest entropy value of 6.80, indicating that this method produced images with the richest information details. Regarding CEIQ, SMO PMD had the lowest value of 3.39, indicating suboptimal enhancement of contrast quality. SMO CLAHE achieved a higher CEIQ value of 3.87, while the combination of SMO PMD-CLAHE delivered the best results with a CEIQ value of 3.97.

The results demonstrate that the combination of SMO PMD-CLAHE delivers the best performance across almost all evaluation metrics. This combination effectively improves illumination, contrast, and image information details. It outperforms both SMO PMD and SMO CLAHE when applied individually. In medical image analysis, particularly for Pap smear images, optimal image quality is crucial for supporting more accurate diagnostic processes. Therefore, using the SMO PMD-CLAHE combination is recommended to enhance overall image quality. This approach can potentially be applied to other scenarios in medical image processing, where improving image quality plays a vital role in supporting clinical decision-making.

## 4. CONCLUSION

This study presents a practical noise-reduction and contrast-enhancement framework for Pap smear images. The proposed method thoroughly evaluates image quality improvement by focusing on clarity, detail preservation, and contrast enhancement. A hybrid PMD-CLAHE method was optimized using the SMO algorithm to overcome the common problems of noise and low contrast in the Pap smear images. The SMO-PMD-CLAHE hybrid method leverages the noise reduction capabilities of the PMD filter while maximizing contrast enhancement through CLAHE. The SMO algorithm consistently provides superior results in optimizing the PMD filter and CLAHE compared with the PSO and POA algorithms. BRISQUE is introduced as a new objective function for PMD filter optimization. BRISQUE performs significantly better than traditional metrics, such as PSNR and SSIM. Similarly, CEIQ is used as a new objective function for CLAHE optimization. CEIQ is a comprehensive assessment of contrast enhancement using a combination of entropy, cross-entropy, and SSIM. The SMO-PMD-CLAHE hybrid approach achieved the highest performance across all evaluated metrics compared with SMO-PMD or SMO-CLAHE. The proposed method, SMO PMD-CLAHE, significantly improved the Pap smear image quality with noise reduction and contrast enhancement.

**ACKNOWLEDGEMENTS**

This research was supported by the Indonesia Endowment Fund for Education (LPDP) through the Beasiswa Pendidikan Indonesia (BPI) scholarship program. BPI ID. 202327091034.

**REFERENCES**

[1] G. Curigliano *et al.*, "Management of cardiac disease in cancer patients throughout oncological treatment: ESMO consensus recommendations," *Annals of Oncology*, vol. 31, no. 2, pp. 171–190, 2020, doi: https://doi.org/10.1016/j.annonc.2019.10.023.
[2] E. R. Putri, A. Zarkasi, P. Prajitno, and D. S. Soejoko, "Artificial neural network for cervical abnormalities detection on computed tomography images," *IAES International Journal of Artificial Intelligence*, vol. 12, no. 1, pp. 171–179, Mar. 2023, doi: 10.11591/ijai.v12.i1.pp171-179.

[3] D. Sambyal and A. Sarwar, "Recent developments in cervical cancer diagnosis using deep learning on whole slide images: An Overview of models, techniques, challenges and future directions," *Micron*, vol. 173, no. April, p. 103520, 2023, doi: 10.1016/j.micron.2023.103520.

[4] A. Dongyao Jia, B. Zhengyi Li, and C. Chuanwang Zhang, "Detection of cervical cancer cells based on strong feature CNN-SVM network," *Neurocomputing*, vol. 411, pp. 112–127, 2020, doi: 10.1016/j.neucom.2020.06.006.

[5] L. Allahqoli et al., "Diagnosis of Cervical Cancer and Pre-Cancerous Lesions by Artificial Intelligence: A Systematic Review," *Diagnostics*, vol. 12, no. 11, pp. 1–32, 2022, doi: 10.3390/diagnostics12112771.

[6] M. Zhao et al., "SEENS: Nuclei segmentation in Pap smear images with selective edge enhancement," *Future Generation Computer Systems*, vol. 114, pp. 185–194, Jan. 2021, doi: 10.1016/j.future.2020.07.045.

[7] A. Khozaimi and W. Firdaus Mahmudy, "New insight in cervical cancer diagnosis using convolution neural network architecture," *IAES International Journal of Artificial Intelligence (IJ-AI)*, vol. 13, no. 3, p. 3092, Sep. 2024, doi: 10.11591/ijai.v13.i3.pp3092-3100.

[8] V. Kamalaveni, R. A. Rajalakshmi, and K. A. Narayanankutty, "Image denoising using variations of Perona-Malik model with different edge stopping functions," *Procedia Comput Sci*, vol. 58, pp. 673–682, 2015, doi: 10.1016/j.procs.2015.08.087.

[9] S. Anam, Z. Fitriah, and N. Shofianah, "Hybrid of the PMD Filter, the K-means clustering method and the level set method for exudates segmentation," in *Proceedings of the International Conference on Mathematics and Islam*, SCITEPRESS - Science and Technology Publications, Jan. 2018, pp. 108–116. doi: 10.5220/0008517901080116.

[10] S. V. Ezhilraman, S. Srinivasan, and G. Suseendran, "Bilateral Perona-malik Diffusion Filtering based Topological Multitude Feature Vector for Breast Cancer Detection," *Journal of Research on the Lepidoptera*, vol. 51, no. 1, pp. 110–128, Feb. 2020, doi: 10.36872/LEPI/V51I1/301010.

[11] M. M. Rahaman et al., "A survey for cervical cytopathology image analysis using deep learning," *IEEE Access*, vol. 8, pp. 61687–61710, 2020, doi: 10.1109/ACCESS.2020.2983186.

[12] C. Tsiotsios and M. Petrou, "On the choice of the parameters for anisotropic diffusion in image processing," *Pattern Recognit*, vol. 46, no. 5, pp. 1369–1381, May 2013, doi: 10.1016/j.patcog.2012.11.012.

[13] A. Jeelani and M. B. Veena, "Hybridization of PSO and Anisotropic Diffusion in Denoising the Images," *Microelectronics, Electromagnetics and Telecommunications. Lecture Notes in Electrical Engineering*, pp. 463–473, 2018, doi: 10.1007/978-981-10-7329-8_47.

[14] E. A. Tjoa, I. P. Yowan Nugraha Suparta, R. Magdalena, and N. Kumalasari CP, "The use of CLAHE for improving an accuracy of CNN architecture for detecting pneumonia," *SHS Web of Conferences*, vol. 139, p. 03026, May 2022, doi: 10.1051/shsconf/202213903026.

[15] M. Hayati et al., "Impact of CLAHE-based image enhancement for diabetic retinopathy classification through deep learning," in *Procedia Computer Science*, Elsevier B.V., 2022, pp. 57–66. doi: 10.1016/j.procs.2022.12.111.

[16] A. Desiani, M. Erwin, B. Suprihatin, S. Yahdin, A. I. Putri, and F. R. Husein, "Bi-path architecture of cnn segmentation and classification method for cervical cancer disorders based on pap-smear images," *IAENG Int J Comput Sci*, vol. 48, no. 3, pp. 1–9, 2021.

[17] R. C. Chen, C. Dewi, Y. C. Zhuang, and J. K. Chen, "Contrast limited adaptive histogram equalization for recognizing road marking at night based on YOLO models," *IEEE Access*, vol. 11, pp. 92926–92942, 2023, doi: 10.1109/ACCESS.2023.3309410.

[18] S. Saifullah and R. Dreżewski, "Modified histogram equalization for improved CNN medical image segmentation," *Procedia Comput Sci*, vol. 225, pp. 3021–3030, 2023, doi: https://doi.org/10.1016/j.procs.2023.10.295.

[19] S. Asghar et al., "Water classification using convolutional neural network," *IEEE Access*, vol. 11, pp. 78601–78612, 2023, doi: 10.1109/ACCESS.2023.3298061.

[20] H. M. Qassim, N. M. Basheer, and M. N. Farhan, "Brightness Preserving Enhancement for Dental Digital X-ray Images Based on Entropy and Histogram Analysis," *Journal of Applied Science and Engineering*, vol. 22, no. 1, pp. 187–194, 2019, doi: 10.6180/jase.201903_22(1).0019.

[21] L. G. More, M. A. Brizuela, H. L. Ayala, D. P. Pinto-Roa, and J. L. V. Noguera, "Parameter tuning of CLAHE based on multi-objective optimization to achieve different contrast levels in medical images," in *2015 IEEE International Conference on Image Processing (ICIP)*, IEEE, Sep. 2015, pp. 4644–4648. doi: 10.1109/ICIP.2015.7351687.

[22] A. Fawzi, A. Achuthan, and B. Belaton, "Adaptive clip limit tile size histogram equalization for non-homogenized intensity images," *IEEE Access*, vol. 9, pp. 164466–164492, 2021, doi: 10.1109/ACCESS.2021.3134170.

[23] U. Kuran and E. C. Kuran, "Parameter selection for CLAHE using multi-objective cuckoo search algorithm for image contrast enhancement," *Intelligent Systems with Applications*, vol. 12, p. 200051, Nov. 2021, doi: 10.1016/j.iswa.2021.200051.

[24] U. Kuran, E. C. Kuran, and M. B. Er, "Parameter Selection of Contrast Limited Adaptive Histogram Equalization Using Multi-Objective Flower Pollination Algorithm," in *ICECENG 2022*, Springer, 2022, pp. 109–123. doi: 10.1007/978-3-031-01984-5_9.

[25] S. Surya and A. Muthukumaravel, "Adaptive Sailfish Optimization-Contrast Limited Adaptive Histogram Equalization (ASFO-CLAHE) for Hyperparameter Tuning in Image Enhancement," in *Computational Intelligence for Clinical Diagnosis*, Springer International Publishing, 2023, pp. 57–76. doi: 10.1007/978-3-031-23683-9_5.

[26] S. R. Borra, N. P. Tejaswini, V. Malathy, B. Magesh Kumar, and M. I. Habelalmateen, "Contrast Limited Adaptive Histogram Equalization based Multi-Objective Improved Cat Swarm Optimization for Image Contrast Enhancement," in *2023 International Conference on Integrated Intelligence and Communication Systems (ICIICS)*, IEEE, Nov. 2023, pp. 1–5. doi: 10.1109/ICIICS59993.2023.10420959.

[27] Y. R. Haddadi, B. Mansouri, and F. Z. I. Khodja, "A novel medical image enhancement algorithm based on CLAHE and pelican optimization," *Multimed Tools Appl*, Apr. 2024, doi: 10.1007/s11042-024-19070-6.

[28] H. Zhu, Y. Wang, Z. Ma, and X. Li, "A Comparative study of swarm intelligence algorithms for UCAV path-planning problems," *Mathematics*, vol. 9, no. 2, pp. 1–31, Jan. 2021, doi: 10.3390/math9020171.

[29] A. Mittal, A. K. Moorthy, and A. C. Bovik, "No-Reference Image Quality Assessment in the Spatial Domain," *IEEE Transactions on Image Processing*, vol. 21, no. 12, pp. 4695–4708, Dec. 2012, doi: 10.1109/TIP.2012.2214050.

[30] J. Yan, J. Li, and X. Fu, "No-Reference Quality Assessment of Contrast-Distorted Images using Contrast Enhancement," *arXiv preprint arXiv*, pp. 1–15, Apr. 2019, doi: https://doi.org/10.48550/arXiv.1904.08879.

[31] R. Kumar and A. K. Bhandari, "Noise reduction deep CNN-based retinal fundus image enhancement using recursive histogram," *Neural Comput Appl*, vol. 36, no. 27, pp. 17221–17243, Sep. 2024, doi: 10.1007/s00521-024-09996-1.

[32] K. P. Win, Y. Kitjaidure, K. Hamamoto, and T. Myo Aung, "Computer-assisted screening for cervical cancer using digital image processing of pap smear images," *Applied Sciences*, vol. 10, no. 5, p. 1800, Mar. 2020, doi: 10.3390/app10051800.

[33] M. E. Plissiti, P. Dimitrakopoulos, G. Sfikas, C. Nikou, O. Krikoni, and A. Charchanti, "Sipakmed: A New Dataset for Feature and Image Based Classification of Normal and Pathological Cervical Cells in Pap Smear Images," in *2018 25th IEEE International Conference on Image Processing (ICIP)*, 2018, pp. 3144–3148. doi: 10.1109/ICIP.2018.8451588.

[34] S. Ray, K. G. Dhal, and P. Kumar Naskar, "Particle swarm optimizer based epithelial layer segmentation in CIElab color space," in *2022 IEEE 7th International Conference on Recent Advances and Innovations in Engineering (ICRAIE)*, IEEE, Dec. 2022, pp. 331–336. doi: 10.1109/ICRAIE56454.2022.10054261.

[35] P. Perona and J. Malik, "Scale-space and edge detection using anisotropic diffusion," *IEEE Trans Pattern Anal Mach Intell*, vol. 12, no. 7, pp. 629–639, Jul. 1990, doi: 10.1109/34.56205.

[36] A. V Nasonov, N. V Mamaev, O. S. Volodina, and A. S. Krylov, "Automatic Choice of Denoising Parameter in PeronaMalik Model," *Lomonosov Moscow State University*, pp. 1–4, 2019.

[37] B. Maiseli, "Nonlinear anisotropic diffusion methods for image denoising problems: Challenges and future research opportunities," *Array*, vol. 17, p. 100265, Mar. 2023, doi: 10.1016/j.array.2022.100265.

[38] S. Muniyappan, A. Allirani, and S. Saraswathi, "A novel approach for image enhancement by using contrast limited adaptive histogram equalization method," *2013 4th International Conference on Computing, Communications and Networking Technologies, ICCCNT 2013*, pp. 1–6, 2013, doi: 10.1109/ICCCNT.2013.6726470.

[39] M. Widyaningsih, T. K. Priyambodo, M. E. Wibowo, and M. Kamal, "Optimization Contrast Enhancement and Noise Reduction for Semantic Segmentation of Oil Palm Aerial Imagery.," *International Journal of Intelligent Engineering & Systems*, vol. 16, no. 1, 2023, doi: 10.22266/ijies2023.0228.51.

[40] J. C. Bansal, H. Sharma, S. S. Jadon, and M. Clerc, "Spider Monkey Optimization algorithm for numerical optimization," *Memet Comput*, vol. 6, no. 1, pp. 31–47, Mar. 2014, doi: 10.1007/s12293-013-0128-0.

[41] V. Kamble and K. M. Bhurchandi, "No-reference image quality assessment algorithms: A survey," *Optik (Stuttg)*, vol. 126, no. 11–12, pp. 1090–1097, Jun. 2015, doi: 10.1016/j.ijleo.2015.02.093.

[42] U. Sara, M. Akter, and M. S. Uddin, "Image Quality Assessment through FSIM, SSIM, MSE and PSNR—A Comparative Study," *Journal of Computer and Communications*, vol. 07, no. 03, pp. 8–18, 2019, doi: 10.4236/jcc.2019.73002.

[43] C. Avatavului and M. Prodan, "Evaluating Image Contrast: A Comprehensive Review and Comparison of Metrics," *Journal of Information Systems & Operations Management*, vol. 17, no. 1, pp. 143–160, 2023.

# BIOGRAPHIES OF AUTHORS

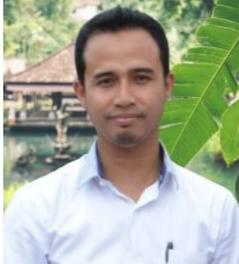 **Ach Khozaimi** is a lecturer in informatics engineering at Trunojoyo University of Madura. He obtained a bachelor's degree in informatics from Trunojoyo University of Madura, Indonesia, and a master's degree in informatics from Institut Teknologi Sepuluh November (ITS), Indonesia. He is pursuing his doctorate in the Department of Mathematics at Brawijaya University, Indonesia. The doctoral program is funded by the Higher Education Financing Centre (BPPT) of the Ministry of Education and Culture Research Technology of the Republic of Indonesia.

**Isnani Darti** stands as the 24th active Professor at the Faculty of Mathematics and Natural Sciences (FMIPA). Her research interests span several captivating domains: Applied Dynamical Systems, Mathematical Biology, Optical Solitons, and Discretization of Continuous Dynamical Systems. Recently, Dr. Isnani Darti achieved the prestigious rank of **Professor** at Brawijaya University. She is the **24th active professor at the Faculty of Mathematics and Natural Sciences (FMIPA) and the 176th active professor** at the university overall. Her professorship adds to the rich legacy of Brawijaya University, where she is the **335th Professor** in its history.

**Syaiful Anam**, Lecturer at the Department of Mathematics, Faculty of Science, Universitas Brawijaya in Malang, Indonesia. Syaiful Anam is passionate about photography and actively involved in education and research. Here's a summary of his educational background: Bachelor's Degree in Mathematics, Universitas Brawijaya, Malang, Indonesia. Master's Degree in Electrical Engineering: Multimedia Intelligent Networks, Institut Teknologi Sepuluh Nopember, Surabaya, Indonesia. Ph.D. in Information and Systems Science - Natural and Mathematical Sciences, Universitas Yamaguchi, Yamaguchi, Japan.

**Wuryansari Muharini Kusumawinahyu**, a lecturer at the Department of Mathematics, Faculty of Science, Universitas Brawijaya in Malang, Indonesia. She is actively involved in teaching and research. Here's a summary of his educational background: Bachelor's Degree (S1), Master's Degree (S2) and Mathematics, Doctoral Degree (S3) from Isntitute Teknologi Bandung, Bandung, Indonesia. Her research interests span various mathematical topics, and she has contributed to several areas, including epidemiology, predator-prey models, and wave dynamics.